\documentclass[aps,prl,showpacs,twocolumn]{revtex4}
\input{tcilatex}

\begin{document}

\title{Interband impact ionization and nonlinear absorption of terahertz
radiations in semiconductor heterostructures\footnote{Published in: J. C. Cao, Physical Review Letters 91, 237401 ( 2003).}}
\author{J. C. Cao}
\affiliation{State Key Laboratory of Functional Materials for Informatics, Shanghai
Institute of Microsystem and Information Technology, Chinese Academy of
Sciences, 865 Changning Road, Shanghai 200050, P. R. China}

\begin{abstract}
We have theoretically investigated nonlinear free-carrier absorption of
terahertz radiation in InAs/AlSb heterojunctions. By considering multiple
photon process and conduction-valence interband impact ionization (II), we
have determined the field and frequency dependent absorption rate. It is
shown that (i) electron-disorder scatterings are important at low to
intermediate field, and (ii) most importantly, the high field absorption is
dominated by II processes. Our theory can satisfactorily explain a long
standing experimental result on the nonlinear absorption in THz regime.
\end{abstract}

\pacs{78.20.Ci, 73.50.Fq, 73.50.Gr, 72.10.Bg}
\maketitle

Many terahertz (THz) related physical phenomena \cite%
{Asmar,BJ1,APL96,SCIENCE,NATURE,CaoPRL}, such as THz-radiation-induced dc
current suppression \cite{Asmar}, multiphoton-assisted resonant tunneling 
\cite{BJ1}, and multiphoton-assisted free-carrier absorption \cite{APL96},
have recently been reported. A few years ago, THz transmissions,
reflections, and dc photoconductivity measurements were made to study
multiphoton process and impact ionization (II) caused by THz radiations in
InAs/AlSb heterojunctions (HJ). These experimental advances motivated new
theoretical studies \cite{CZhang,WXU,LeiTHzJAP,CaoJAP,CaoPRB4} of THz
physics. However, many interesting phenomena observed experimentally
remained unexplained due to lack of well controlled theoretical models in
the nonlinear regime. For example, the ac field dependent absorption
coefficient \cite{APL96} increases with the field slowly at low field
intensity, then increases rapidly at intermediate field intensities and
levels off at high intensities. To date, this behavior is still not fully
understood even within the picture of multiphoton absorption mechanism.

In this letter, we propose to explain this interesting absorption process
for the first time by studying the multiphoton-assisted free-carrier
absorption and electron-hole (e-h) generation in THz-driven InAs/AlSb HJ's.
Most existing theoretical work emphasize the importance of the
electron-impurity (e-i) \cite{CZhang} or the electron-phonon (e-p)
interaction \cite{WXU,LeiTHzJAP} in nonlinear absorption. However, both
these two scattering mechanisms lead to a decreasing absorption in the high
field limit and thus unable to explain the observed phenomena. The parameter
that controls the absorption is the electro-optic coupling factor, $\mathbf{r%
}_{\omega }=e\mathbf{E}_{\mathrm{ac}}/(m\omega ^{2})$, here $e$ is the
carrier charge, $m$ is the effective electron mass, $E_{\mathrm{ac}}$ is the
amplitude, and $\omega =2\pi f_{\mathrm{ac}}$ is the angular frequency with $%
f_{\mathrm{ac}}$ the radiation frequency. In what follows we shall show that
the dominant contribution to the nonlinear absorption in the intermediate to
high field region is from the interband II processes. The calculated
free-carrier absorption percentages of THz radiations in InAs/AlSb HJ are in
excellent agreement with the experimental data \cite{APL96}.

Consider a semiconductor HJ with a two-dimensional (2D) energy-wavevector
relation $\varepsilon _{s}(\mathbf{k}_{\parallel })$, where $s$ is the
subband index. According to the balance-equation theory, when a uniform dc
(or slowly varying) electric field $\mathbf{E}_{0}$ and a uniform sinusoidal
radiation field, $\mathbf{E}(t)=\mathbf{E}_{0}+\mathbf{E}_{\mathrm{ac}}\sin
(2\pi f_{\mathrm{ac}}t),$ are applied in the direction parallel to HJ
interface, the transport state can be described by the following equations
for the force, the energy, and the carrier number balance \cite%
{LeiTHzJAP,CaoJAP,LeiPRL},%
\begin{equation}
\frac{\mathrm{d}\mathbf{v}}{\mathrm{d}t}=e\mathbf{E}_{0}\cdot \mathcal{K}+%
\mathbf{A}_{\mathrm{ei}}+\mathbf{A}_{\mathrm{ep}}+\mathbf{A}_{\mathrm{II}}-g%
\mathbf{v},  \label{vdeqn}
\end{equation}%
\begin{equation}
\frac{\mathrm{d}h_{e}}{\mathrm{d}t}=e\mathbf{E}_{0}\mathbf{\cdot v-}W_{%
\mathrm{ep}}-W_{\mathrm{II}}-gh_{e}+S_{\mathrm{i}}+S_{\mathrm{p}}+S_{\mathrm{%
II}},  \label{egyeqn}
\end{equation}%
\begin{equation}
\frac{\mathrm{d}N}{\mathrm{d}t}=gN,  \label{nseqn}
\end{equation}%
where $\mathbf{v}$ is the average electron velocity, and $\mathcal{K}$ is
the inverse effective mass tensor. The electron sheet density is $N=2\sum_{s,%
\mathbf{k}_{\parallel }}f\left[ \left( \varepsilon _{s}(\mathbf{k}%
_{\parallel })-\mu \right) /T_{e}\right] $ with $T_{e}$ the electron
temperature and $\mu $ the electron chemical potential. $f(x)=1/\left[ \exp
\left( x\right) +1\right] $ is the Fermi distribution function. $g$ is the
net electron-hole generation rate by balancing II generation and Auger
recombination. $\mathbf{A}_{\mathrm{ei}},$ $\mathbf{A}_{\mathrm{ep}},$ and $%
\mathbf{A}_{\mathrm{II}}$ are the frictional accelerations respectively due
to e-i scattering, e-p scattering, and interband II process. $W_{\mathrm{ep}%
} $ and $W_{\mathrm{II}}$ are the energy-loss rates respectively due to e-p
scattering and interband II process. $S_{\mathrm{i}}$, $S_{\mathrm{p}}$ and $%
S_{\mathrm{II}}$ are the energy-gain rates of the electron system from the
radiation field through the multiphoton process ($n=\pm 1,\pm 2,...$) in
association with e-i interaction, e-p interaction, and II process,
respectively. The expression of $S_{\mathrm{II}}$ is as follow, 
%\begin{widetext}
\begin{eqnarray}
S_{\mathrm{II}} &=&\frac{2}{N}\sum_{s^{\prime },s,\mathbf{k}_{\parallel }%
\mathbf{,q}_{\parallel }}\left\vert \widetilde{M}_{s^{\prime },s}^{\mathrm{II%
}}\left( \mathbf{q}_{\parallel }\right) \right\vert ^{2}\sum_{n=-\infty
}^{\infty }J_{n}^{2}(\mathbf{q}_{\parallel }\cdot \mathbf{r}_{\omega
})n\omega  \nonumber \\
&\times &\Pi _{2}(s^{\prime },s,\mathbf{q}_{\parallel },\omega _{3})\left[
f\left( \frac{\xi _{\mathbf{k}_{\parallel }\mathbf{-q}_{\parallel }}}{T_{e}}%
\right) +n\left( \frac{\omega _{3}}{T_{e}}\right) \right]  \nonumber \\
&\times &\left[ f\left( \frac{\xi _{\mathbf{k}_{\parallel }}}{T}\right)
-f\left( \frac{\omega _{1}+mv^{2}/2+\varepsilon _{k}^{h}+\mu }{T_{e}}\right) %
\right] ,  \label{SII}
\end{eqnarray}%
%\end{widetext}
where $T$ is the lattice temperature, $\mathbf{q}_{\parallel }=(q_{x},q_{y})$
is the 2D phonon wave vector, $\omega _{1}=k_{x}v,$ $\omega _{2}=\omega
_{1}+mv^{2}/2+\varepsilon _{\mathbf{k}_{\parallel }}^{h}+\varepsilon _{s,%
\mathbf{k}_{\parallel }-\mathbf{q}_{\parallel }},$ $\omega _{3}=\omega
_{2}-n\omega ,$ $\xi _{\mathbf{k}_{\parallel }}=\varepsilon _{s,\mathbf{k}%
_{\parallel }}-\mu ,$ $\xi _{\mathbf{k}_{\parallel }}^{h}=\varepsilon _{%
\mathbf{k}_{\parallel }}^{h}-\mu ^{h},$ $\varepsilon _{\mathbf{k}_{\parallel
}}^{h}=E_{g}+\mathbf{k}_{\parallel }^{2}/(2m_{h})$ is the hole dispersion
with $m_{h}$ the effective hole mass. $n(x)=1/[\exp (x)-1]$ is the Bose
function, and $E_{g}$ is the band gap. $\Pi _{2}(s^{\prime },s,\mathbf{q}%
_{\parallel },\Omega )$ is the imaginary part of electron-electron
correlation function \cite{LeiTHzJAP}. $J_{n}(x)$ is the $n$th-order Bessel
function. $\widetilde{M}_{s^{\prime },s}^{\mathrm{II}}(\mathbf{q}_{\parallel
})$ is the Fourier representation of the band-band Coulomb interaction
matrix element \cite{Ando} for the II and Auger processes in the 2D
semiconductor system. The total energy-gain rate is defined by 
\begin{equation}
S=S_{\mathrm{i}}+S_{\mathrm{p}}+S_{\mathrm{II}}.  \label{totS}
\end{equation}%
It is useful to write $S$ as a sum of all orders of $n$-photon
contributions, $S=\sum_{n=1}^{\infty }S_{n},$ where $S_{n}$ is the total
contribution from terms having index $n$ and $-n$. We introduce the quantity 
$\alpha $ which is the ratio of the electromagnetic (EM) energy loss through
the 2D sheet to the energy of the incident EM wave. $\alpha $ is a measure
of the absorption of the radiation which can be written as, 
\begin{equation}
\alpha =\frac{2NS}{\sqrt{\kappa }\varepsilon _{0}cE_{\mathrm{ac}}^{2}},
\label{alp}
\end{equation}%
where $c$ is the light speed in vacuum and $\kappa $ is the dielectric
constant of the semiconductor. To see the role of individual multiphoton
processes, we define $\alpha _{n}=2NS_{n}/(\sqrt{\kappa }\varepsilon _{0}cE_{%
\mathrm{ac}}^{2})$ and write $\alpha =\sum_{n=1}^{\infty }\alpha _{n}$.

In the steady-state calculations for THz-driven InAs/AlSb HJ's, we consider
the electron-acoustic-phonon scattering (via the deformation potential and
the piezoelectric couplings), electron-polar-optical-phonon scattering (via
the Fr\"{o}hlich coupling), and elastic scattering both from the remote
charged impurities and from the background impurities. For InAs, the
Kane-type nonparabolic factor is $2.73$ eV$^{-1}$ and $E_{g}=0.22$ eV. The
three lowest electron subbands ($\varepsilon _{0}=0,$ $\varepsilon _{1}=35$
meV, and $\varepsilon _{2}=200$ meV) are included. Up to 10 hole subbands
are taken into account. We set the sheet density of InAs well $N_{0}=5\times
10^{12}$ cm$^{-2}$, depletion layer charge density $N_{\mathrm{dep}}=5\times
10^{10}$ cm$^{-2}$, background impurity $n_{I}=6.86\times 10^{15}$cm$^{-3}$,
and remote impurities in AlSb barrier $N_{I}=1.53\times 10^{11}$cm$^{-2}$
located at a distance of $l=10$ nm\ from the HJ interface. In the whole
paper, we set $T=300$ K, and the dc field is assumed to be $E_{0}=1.5$ V/m,
same as that used in the experiments \cite{APL96}. The orders of the Bessel
functions, $n$, in Eq. (\ref{SII}) is set to be large enough so that the
integral is convergent within the accuracy of $10^{-4}$.

In Fig. \ref{fig1}(a) we show the e-h pair generation (solid lines) vs $E_{%
\mathrm{ac}}$ with the II process at $f_{\mathrm{ac}}=0.64$, 1, and 1.7 THz,
respectively. The dash line in Fig. \ref{fig1}(a) shows no changes of the
carrier number when the II process is excluded. The generation rate is
essentially zero at low fields and high frequencies. If we define that the
onset of the II process occurs at the threshold field $E_{\mathrm{c}}(f_{%
\mathrm{ac}})$. The calculations indicate that $E_{\mathrm{c}}(f_{\mathrm{ac}%
})/f_{\mathrm{ac}}$ is nearly constant for all frequencies (see the inset of
Fig. \ref{fig1}(a), $E_{\mathrm{c}}(f_{\mathrm{ac}})/f_{\mathrm{ac}}\simeq
3.59$). The physical meaning of this linear dependence is such that for II
process to take place, $r_{\omega }$ times the photon energy $\hbar \omega $
must exceed a minimum value, $r_{\omega }\hbar \omega =$ constant. At the
high field $N$ depends on $E_{\mathrm{ac}}$ exponentially. We shall show
below that the exponentially growing generation rate plays a crucial role in
the total absorption percentage. The Fig. \ref{fig1}(b) shows the effect of
II processes on the electron temperature. While the mechanisms of the
temperature increase with the coupling parameter $r_{\omega }$ is generally
understood \cite{WXU2}, our results show a surprisingly large contribution
to $T_{e}$ from the interband II processes. The interband II processes
account for about 50 percent of $T_{e}$. It is interesting to note that II
processes lead to a net increase of $T_{e}$. In many cases, II processes act
as a cooling mechanism. For example, for a 3D system under a dc field, $T_{e}
$ with II process is lower than that without II process. In this case the
electron heating is only due to electron accelerating and scattering under
the dc field. For the present system under an ac field, the II processes
still act as a cooling mechanism. However, large number of electrons created
in II processes immediately causes a rapid increase in photon absorption due
to e-i, e-p, and more importantly, e-h scattering. This absorption is
further enhanced by nonlinear multiphoton absorption under an ac field. This
subsequent electron scattering and absorption of photon energy lead to an
increase of $T_{e}$ which exceeds the initial reduction of $T_{e}$ due to II
processes. As a result, $T_{e}$ is higher with II processes.

\FRAME{ftbpFU}{8.3472cm}{4.3735cm}{0pt}{\Qcb{(a) Electron-hole pair
generation $N/N_{0}$ and (b) electron temperature $T_{e}/T$ vs $E_{\mathrm{ac%
}}$ with and without II process in InAs/AlSb HJ at $f_{\mathrm{ac}}=0.64$,
1, and 1.7 THz, respectively.}}{\Qlb{fig1}}{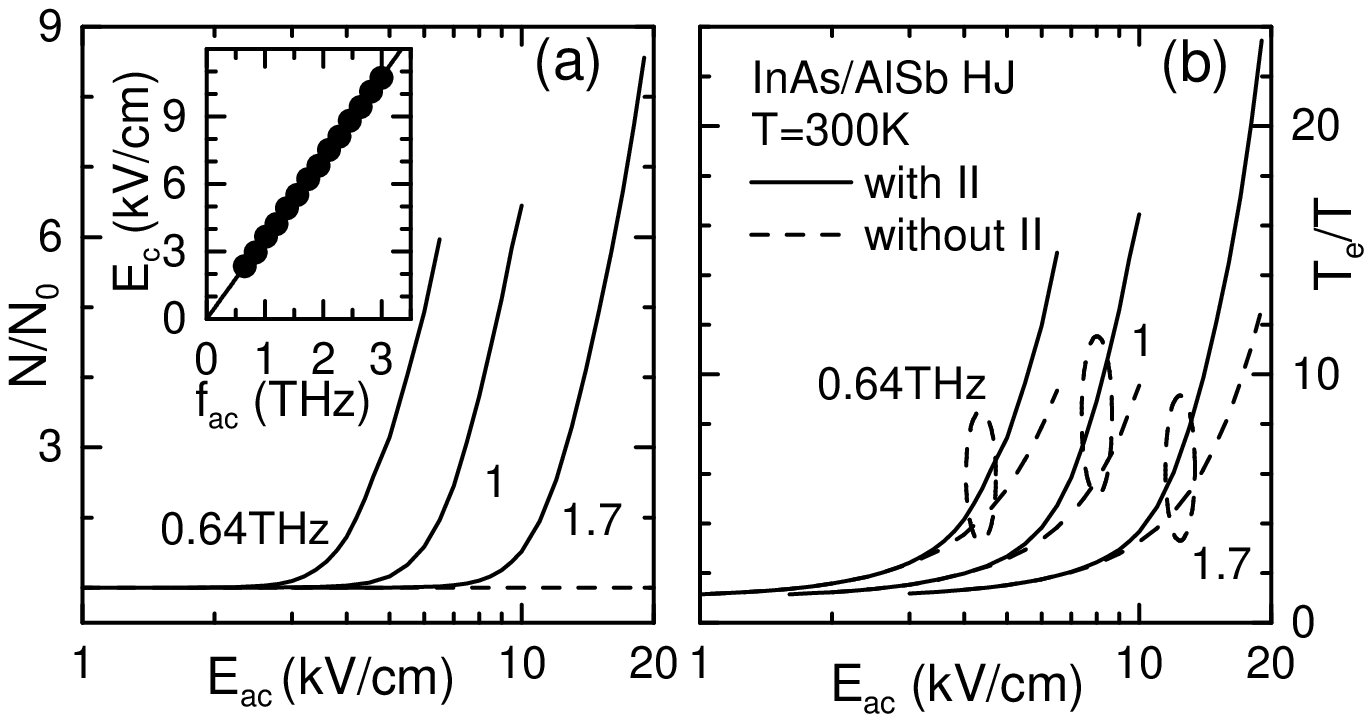}{\special{language
"Scientific Word";type "GRAPHIC";maintain-aspect-ratio TRUE;display
"USEDEF";valid_file "F";width 8.3472cm;height 4.3735cm;depth
0pt;original-width 5.8072in;original-height 2.8271in;cropleft "0";croptop
"1";cropright "1";cropbottom "0";filename 'fig1.eps';file-properties
"XNPEU";}}

In Fig. \ref{fig2}(a) we show the calculated total absorption percentage $%
\alpha $ vs $E_{\mathrm{ac}}$ at $f_{\mathrm{ac}}=0.64$, 1, and 1.7 THz,
respectively. The solid circles are the experimental results for $f_{\mathrm{%
ac}}=0.64$ THz from Ref. \onlinecite{APL96}, where $\alpha =1-T_{r}-R$ with $%
T_{r}$ the transmission and $R$ the reflection. Excellent agreement is
obtained between the calculated results and the experimental data. It's seen
from Fig. \ref{fig2}(a) that, lower frequency and/or higher radiation
intensity lead to more absorption. For $f_{\mathrm{ac}}=0.64$ THz, the
absorption rate increases slowly from low electric field to about $E_{%
\mathrm{ac}}=3$ kV/cm, then goes fast before reaching its saturation value.
The significance of II process can be clearly seen in the measurements and
in the current theory. Without the II process, the multiphoton absorption is
due to e-i scattering and e-p scattering. The absorption rate due to these
mechanism only increases very slowly at low $E_{\mathrm{ac}}$. In Fig. \ref%
{fig2}(b) we show the phonon-induced ($S_{\mathrm{p}}$), II-induced ($S_{%
\mathrm{II}}$), and the total energy gains ($S_{\mathrm{p}}+S_{\mathrm{II}}$%
) of the electron system from the radiation fields. At intermediate to high $%
E_{\mathrm{ac}}$, both e-i and e-p scatterings are actually suppressed by
the strong electro-optic coupling. As a result one observes a rapidly
decreasing absorption in the high field limit. As we have seen in Fig. \ref%
{fig1}, the effect of strong field on the II process is quite opposite to
that of electron disorder scattering because the e-h pair generation is a
fundamentally different process than the scattering process. Scattering
depends on the ability of electrons to simultaneously exchange momentum and
energy with the scatters and photon. Strong coupling with photons reduce the
probability of momentum exchange with the impurities or phonons. This
momentum exchange is not required in an e-h pair generation process.
Annihilation of multiple photon in a generation process leads to a strong
total absorption percentage. The very rapid increase of II process at high
field [see Fig. \ref{fig1}(a)] is largely offset by the fast decreasing of
impurity and phonon scattering. The resulting total absorption percentage
increase modestly in the high field limit. The curve at $f_{\mathrm{ac}}=0.64
$ THz can be directly compared to experimental result. At this frequency, II
mechanism begins to make contribution to the total absorption after about $%
E_{\mathrm{ac}}=3$ kV/cm. The carrier number increases with increasing $E_{%
\mathrm{ac}}$. For $f_{\mathrm{ac}}=0.64$ THz, when $E_{\mathrm{ac}}=6.5$
kV/cm, the total electron number $N$ increases up to about six times the
initial number $N_{0}$ due to the II process. The agreement between the
present theory and the experiment is excellent. In the inset of Fig. \ref%
{fig2}(b) we show the $n$-photon absorption percentage $\alpha _{n}$ vs the
order $n$ of the Bessel function at $f_{\mathrm{ac}}=1.7$ THz. $E_{\mathrm{ac%
}}$ is changed from 2 to 6 kV/cm with a step of 0.5 kV/cm. For each $E_{%
\mathrm{ac}}$ there is a maximal of $\alpha _{n}$, and the position of the
peaks shifts to higher order $n$ with increasing $E_{\mathrm{ac}}$. It means
that higher order multiphoton processes play more important role on the
total absorption with increasing $E_{\mathrm{ac}}$. Generally, it can be
seen that the role of THz radiations with larger $E_{\mathrm{ac}}$ and lower 
$f_{\mathrm{ac}}$ on carrier transport increases with increasing the
electric-optic coupling factor $r_{\omega }$.

\FRAME{fbFU}{8.3274cm}{10.0122cm}{0pt}{\Qcb{(a) Absorption percentage $%
\protect\alpha $ vs $E_{\mathrm{ac}}$ at $f_{\mathrm{ac}}=$0.64, 1, and 1.7
THz, respectively. The circles are the experimental results from Ref. 3. (b)
The corresponding phonon-induced ($S_{\mathrm{p}}$), II-induced ($S_{\mathrm{%
II}}$), and total energy gains ($S_{\mathrm{p}}+S_{\mathrm{II}}$) vs $E_{%
\mathrm{ac}}$. The inset shows $n$-photon absorption percentage $\protect%
\alpha _{n}$ vs the order $n$ of the Bessel function at $f_{\mathrm{ac}}=1.7$
THz.}}{\Qlb{fig2}}{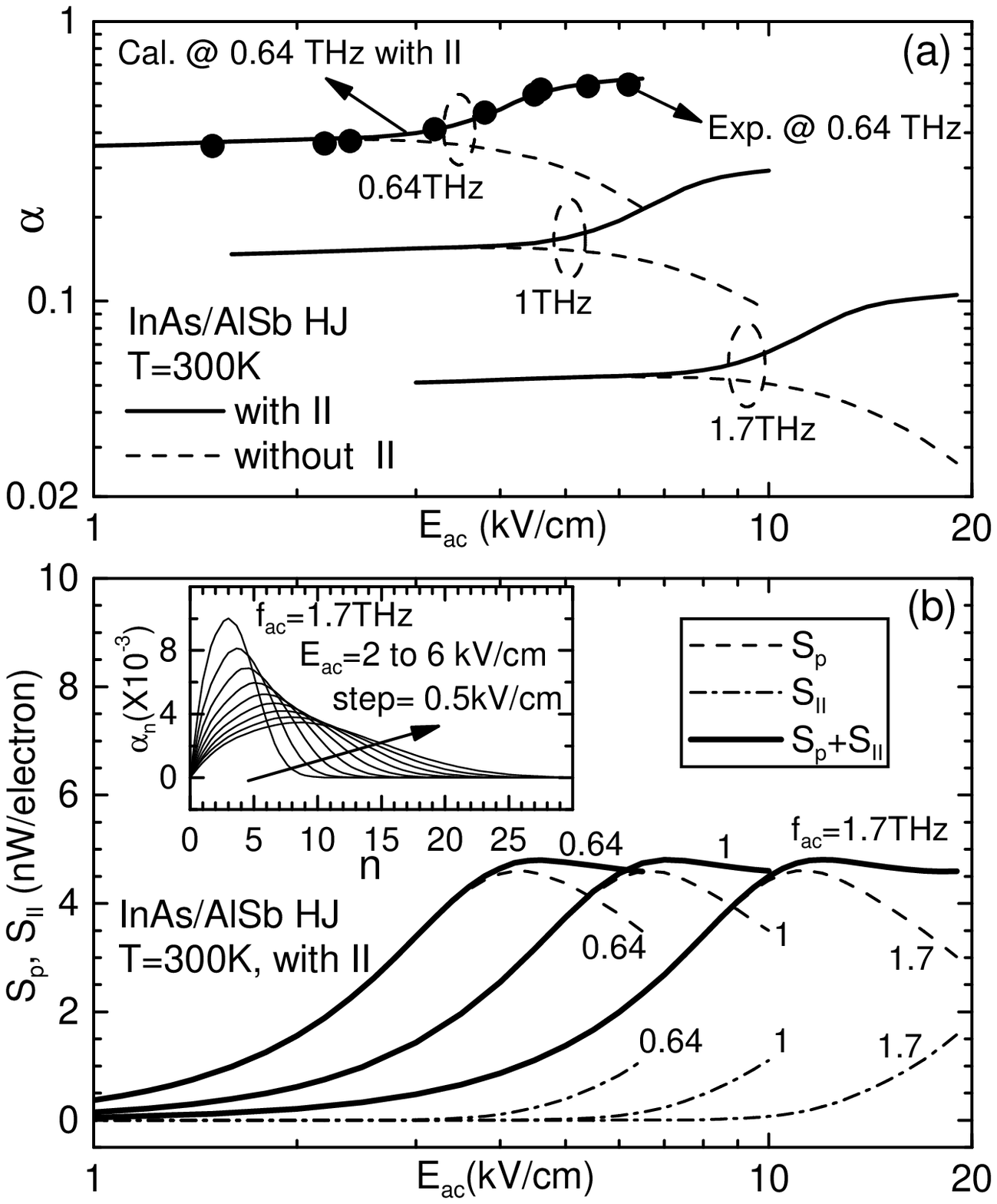}{\special{language "Scientific Word";type
"GRAPHIC";maintain-aspect-ratio TRUE;display "USEDEF";valid_file "F";width
8.3274cm;height 10.0122cm;depth 0pt;original-width 5.3956in;original-height
6.5138in;cropleft "0";croptop "1";cropright "1";cropbottom "0";filename
'fig2.eps';file-properties "XNPEU";}}

\FRAME{ftbpFU}{8.3186cm}{6.5196cm}{0pt}{\Qcb{Calculated total absorption
percentage $\protect\alpha $ vs wavelength $\protect\lambda $ at $E_{\mathrm{%
ac}}=2$ kV/cm (triangles), 5.8 kV/cm (circles), and 9 kV/cm (squares),
respectively. They are fitted by the function (lines): $\protect\alpha (%
\protect\lambda )=\protect\alpha _{0}\protect\lambda ^{b}$. The inset shows
the first five order multiphoton contribution at $E_{\mathrm{ac}}=2$ kV/cm.}%
}{\Qlb{fig3}}{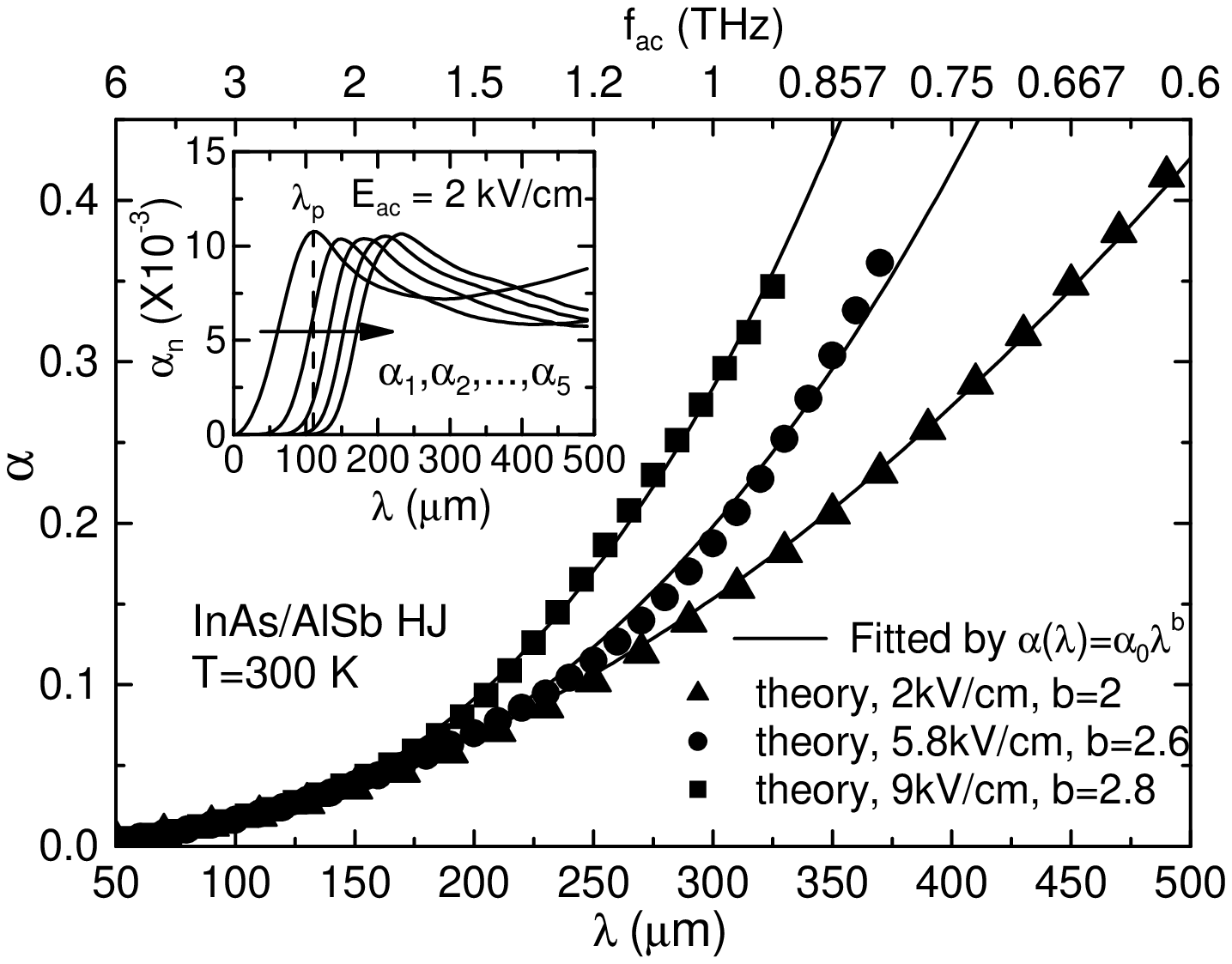}{\special{language "Scientific Word";type
"GRAPHIC";maintain-aspect-ratio TRUE;display "USEDEF";valid_file "F";width
8.3186cm;height 6.5196cm;depth 0pt;original-width 5.7398in;original-height
4.5671in;cropleft "0";croptop "1";cropright "1";cropbottom "0";filename
'fig3.eps';file-properties "XNPEU";}}

In Fig.\thinspace \ref{fig3} we show the total absorption percentage $\alpha 
$ vs the wavelength $\lambda $ at $E_{\mathrm{ac}}=2$ kV/cm (triangles), 5.8
kV/cm (circles), and 9 kV/cm (squares), respectively. The absorption
percentage $\alpha $ increases with increasing $E_{\mathrm{ac}}$ and/or
increasing $\lambda $. The relation between $\alpha $ and $\lambda $ can be
fitted by the function (see the lines in Fig. \ref{fig3}): $\alpha (\lambda
)=\alpha _{0}\lambda ^{b}$. When $E_{\mathrm{ac}}=2$ kV/cm the fitted
function is $\alpha (\lambda )=1.70\times 10^{-6}\lambda ^{2}$, which is the
classical absorption-wavelength relation. When $E_{\mathrm{ac}}$ increases,
however, the present nonlinear theory predicts that the square relation will
be broken. For $E_{\mathrm{ac}}=$ 5.8 kV/cm we have $(\alpha
_{0},b)=(7.19\times 10^{-8},2.6)$, and for $E_{\mathrm{ac}}=9$ kV/cm we have 
$(\alpha _{0},b)=(3.31\times 10^{-8},2.8)$. The physical reason is the
exponential growth of the e-h pair and the more importance of the
multiphoton process with increasing $r_{\omega }$. In the inset of Fig. \ref%
{fig3}, we show the first five order multiphoton contribution $\alpha _{n}$ (%
$n=1,2,...,5$) vs $\lambda $ at $E_{\mathrm{ac}}=2$ kV/cm. It can be seen
that, for each order of multiphoton channels there is a critical wavelength $%
\lambda _{p}$ at which the multiphoton process makes maximal contribution to
the total absorption. With increasing $n$, the position of $\lambda _{p}$
shifts to longer wavelength. The physical origin of the critical wavelength
at a given photon number is that, for a given $n$, the strength of the
scattering matrix of the $n$-photon assisted process decreases almost like $%
\omega ^{-4n}$ for large $\omega $, while the energy absorpted from the
radiation field during each $n$-photon process is proportional to $n\omega $
[see Eq. (\ref{SII})]. There must be a critical wavelength $\lambda _{p}$ at
which the energy absorption reaches a maximum.

Several physical mechanisms employed in our theoretical model and numerical
calculation need to be further discussed. (1) We have neglected the
absorption of bulk electrons in the substrate because the total absorption
is dominated by the 2D channel. The Coulomb interaction matrix for the 2D
channel is much stronger than the corresponding interaction matrix of 3D
system except in the small wavevector regime. On the other hand the
electron-laser coupling $\mathbf{q}_{\parallel }\cdot \mathbf{r}_{\omega }$
[see Eq. (\ref{SII})] is only important at large $q_{\parallel }$ regime. As
a result, the THz-induced II in the bulk plays a negligible role in the
present system. (2) Although a large number of holes are being created in II
processes, the hole contribution to the total absorption is ignored in our
model. The hole effective mass ($m_{h}=0.94m_{0}$) in our system is about 25
times heavier than that of the electron ($m=0.038m_{0}$), here $m_{0}$ is
the mass of the free electron. From Eq. (4), one can see that $S_{\mathrm{II}%
}$ is mainly determined by $J_{n}^{2}(\mathbf{q}_{\parallel }\cdot \mathbf{r}%
_{\omega })$. For a small $r_{\omega },$ i.e., larger mass $m$ or higher
frequency $\omega $, we have $J_{n}^{2}(\mathbf{q}_{\parallel }\cdot \mathbf{%
r}_{\omega })\propto m^{-2|n|}$ for $|n|>0$. For $n=0$, we can directly see
from Eq. (4) and the inset of Fig. \ref{fig2}(b) that the term does not
contribute to the total absorption. When the electron mass is replaced by
the hole mass in $J_{n}^{2}(\mathbf{q}_{\parallel }\cdot \mathbf{r}_{\omega
})$ (for $|n|>0$), this quantity will be reduced by a factor of around $%
25^{-2}=0.0016$. This makes the hole contribution at least two to three
orders of magnitude smaller than that of the electrons. (3) We have used a
strict 2D model for charged carriers and neglected the effect of real space
transfer (RST) to bulk electrons. For InAs/AlSb HJ system, the confining
potential is $\Delta E_{c}=1350$ meV \cite{NakaAPL89}. This is about 15666
K. The maximum $T_{e}$ in our calculation is around 20 times the lattice
temperature (about 6000 K). Since the case $f_{\mathrm{ac}}=0.64$ THz
represents our central result, we estimate the RST effect. At $E_{\mathrm{ac}%
}=6$ kV/cm, $T_{e}$ is about $12T$ or 3600 K, which is less than a quarter
of the confining potential. If the RST is due to over-the-barrier
activation, the probability of this transfer is $\sim \exp (-\Delta
E_{c}/T_{e})\approx 0.0129$, which is less than 1.5 percent. On the other
hand, in terms of photon energy of 1 THz (about 4 meV), the barrier height
is about 330 times of photon energy. This makes the RST by gaining enough
energy from the THz field also negligible. Actually, it is clear from the
inset of Fig. \ref{fig2}(b) that the probability of absorbing more than 30
photons is practically zero. Therefore we neglect the RST effect.

In conclusion, we have shown that II process plays an important role in the
free-carrier absorption of THz radiations in InAs/AlSb HJs. Our results can
explain a long standing experimental result on the free carrier absorption
in the THz regime.

This work was supported by the key National Natural Science Foundation of
China and the Funds with Nos. 2001CCA02800G, 20000683, and 011661075. The
author thanks Prof. X. L. Lei for many helpful discussions.


\begin{thebibliography}{99}
\bibitem{Asmar} N. G. Asmar \emph{et al.}, Phys. Rev. B \textbf{51}, 18041
(1995).

\bibitem{BJ1} B. J. Keay \emph{et al.}, Phys. Rev. Lett. \textbf{75}, 4098
(1995).

\bibitem{APL96} A. G. Markelz \emph{et al.}, Appl. Phys. Lett. \textbf{69},
3975 (1996).

\bibitem{SCIENCE} R. Paiella \emph{et al.}, Science \textbf{290}, 1739
(2000).

\bibitem{NATURE} R. K\"{o}hler \emph{et al.}, Nature \textbf{417}, 156
(2002).

\bibitem{CaoPRL} H. C. Liu \emph{et al.}, Phys. Rev. Lett. \textbf{90},
077402 (2003).

\bibitem{CZhang} C. Zhang, Phys. Rev. B \textbf{65}, 153107 (2002).

\bibitem{WXU} W. Xu and C. Zhang, Phys. Rev. B \textbf{54}, 4907 (1996).

\bibitem{LeiTHzJAP} X. L. Lei, J. Appl. Phys. \textbf{84}, 1396 (1998).

\bibitem{CaoJAP} J. C. Cao \emph{et al.}, J. Appl. Phys. \textbf{87}, 2867
(2000).

\bibitem{CaoPRB4} J. C. Cao \emph{et al.}, Phys. Rev. B \textbf{63}, 115308
(2001).

\bibitem{LeiPRL} X. L. Lei \emph{et al.}, Phys. Rev. Lett. \textbf{66}, 3277
(1991).

\bibitem{Ando} T. Ando \emph{et al.}, Rev. Mod. Phys. \textbf{54}, 437
(1982).

\bibitem{WXU2} W. Xu and C. Zhang, Phys. Rev. B \textbf{55}, 5259 (1997).

\bibitem{NakaAPL89} A. Nakagawa \emph{et al.}, Appl. Phys. Lett. \textbf{54}%
, 1893 (1989).
\end{thebibliography}
\end{document}